\documentclass[12pt,preprint]{aastex}

\newcommand{\mmhg}{\ensuremath{\,{\rm mm\,Hg}}} 
\newcommand{\cmsq}{\ensuremath{{\rm cm}^2}}	
\newcommand{\per}[1]{\,#1\ensuremath{^{-1}}}    
\newcommand{\percent}{\,\%}			
\newcommand{\cmpmolec}{\ensuremath{\,{\rm cm\ molec^{-1}}}}	
\newcommand{\ee}[1]{\ensuremath{\times10^{#1}}} 
\newcommand{\wav}{\ensuremath{\lambda}} 
\newcommand{\nyu}{\ensuremath{\tilde{\nu}}}
\newcommand{\ammonia}{\ensuremath{{\rm NH_3}}}
\newcommand{\com}{\ensuremath{{\rm CO}}}
\newcommand{\cod}{\ensuremath{{\rm CO_2}}}
\newcommand{\methane}{\ensuremath{{\rm CH_4}}}
\newcommand{\methanol}{\ensuremath{{\rm CH_3OH}}}
\newcommand{\suboxide}{\ensuremath{{\rm C_3O_2}}}
\newcommand{\thcom}{\ensuremath{\rm ^{13}}\com}
\newcommand{\twcom}{\ensuremath{\rm ^{12}}\com}
\newcommand{\water}{\ensuremath{\rm H_2O}}			
\newcommand{\mic}{\ensuremath{\,\mu{\rm m}}}	
\newcommand{\wvn}{\ensuremath{\,{\rm cm}^{-1}}} 
\newcommand{\formaldehyde}{\ensuremath{{\rm H_2CO}}}
\newcommand{\percmsq}{\ensuremath{\,{\rm cm}^{-2}}}	
\newcommand{\K}{\ensuremath{\,{\rm K}}}		
\newcommand{\ns}{\ensuremath{\!}} 

\shorttitle{Near-IR Band Strengths of Astrophysical Ices}
\shortauthors{Gerakines et al.}

\begin{document}

\slugcomment{To appear in {\it the Astrophysical Journal}, 20 February 2005}

\title{The Strengths of Near-Infrared Absorption Features Relevant to
Interstellar and Planetary Ices}

\author{P.~A. Gerakines, J.~J. Bray, A. Davis, and C.~R. Richey}

\affil{Astro- and Solar-System Physics Program, Department of Physics,
University of Alabama at Birmingham, Birmingham, AL 35294-1170}

\email{gerak@uab.edu}

\begin{abstract}
The abundances of ices in planetary environments have historically
been obtained through measurements of near-infrared absorption
features (\wav\ = 1.0--2.5\mic), and near-IR transmission measurements
of materials present in the interstellar medium are becoming more
common.  For transmission measurements, the band strength (or
absorption intensity) of an absorption feature must be known in order
to determine the column density of an ice component.  In the
experiments presented here, we have measured the band strengths of the
near-IR absorption features for several molecules relevant to the
study of interstellar icy grain mantles and icy planetary bodies:
\com\ (carbon monoxide), \cod\ (carbon dioxide), \suboxide\ (carbon
suboxide), \methane\ (methane), \water\ (water), \methanol\
(methanol), and \ammonia\ (ammonia).  During a vacuum deposition, the
sizes of the near-IR features were correlated with that of a studied
mid-IR feature whose strength is well known from previous ice studies.
These data may be used to determine ice abundances from observed
near-IR spectra of interstellar and planetary materials or to predict
the sizes of near-IR features in spectral searches for these molecules
in astrophysical environments.
\end{abstract}


\section{Introduction}

Observations in the mid-infrared spectral region (\nyu\ =
4000-400\wvn; \wav\ = 2.5-25\mic) have led to the identification of
various molecules in the icy mantles that coat the dust grains present
in the interstellar medium (ISM).  These include firm identifications
of \water, \com, \cod, OCN$^-$ (cyanate ion), \methanol, \methane, and
OCS, some reasonable but more tentative identifications of HCOOH
(formic acid), HCOO$^-$ (formate ion), \formaldehyde\ (formaldehyde),
and \ammonia, and the absorption features of some uncharacterized
species such as large organics, or molecules possibly related to
ammonium \citep[][and references
therein]{ehrenfreund00,whittet03,gibb04}.  Observations of spectral
features located in the near-infrared spectral region (\nyu\ =
12500-4000\wvn; \wav\ = 0.8-2.5\mic) have been widely used to identify
ices in the outer Solar System.  Molecules such as \water, \methane,
\com, \cod, and N$_2$ have been identified on planetary bodies such as
Pluto, Charon, Triton, and Europa (see a recent review by
\citealt{roush01} and observations by, e.g., \citealt{cruikshank98},
\citealt{quirico99}, \citealt{buie00}, or \citealt{grundy03}).
Cometary nuclei are presumed to have a composition similar to these
objects \citep[e.g.,][]{irvine00}.

In recent years, observations of the ISM have begun to extend into the
near-IR spectral region as well. Observations of interstellar material
have been made with instruments such as SpeX on the Infrared Telescope
Facility (IRTF) telescope \citep{vacca04} and other instruments
(e.g., \citealt{murakawa00}; \citealt{taban03}).  The instruments designed
for the future James Webb Space Teleacope (JWST) call for an increase
in fundamental near-IR data.  The JWST Near Infrared Spectrometer
(NIRSpec) will perform observations from 0.6 to 5\mic\ at spectral
resolutions from 100 to 3000, with the primary objective of studying
the processes of star formation in the ISM of external galaxies
\citep{rauscher04}.  Calculations of ice abundances in star-forming
regions could be accomplished with near-IR data such as ours.

The relative abundances of ice components can provide key pieces of
information about the physical and chemical history of the dense cloud
in which the ice species reside.  Abundances are obtained by observing
the absorption features of a molecule arising from its allowed
vibrational modes.  For unsaturated features with small peak optical
depths, the integrated area of an interstellar ice absorption feature
scales with the molecule's abundance by a constant factor.  This
factor is called the strength of the absorption feature, and it is an
intrinsic physical parameter of its carrier molecule.  As has been
shown in many previous studies
\citep[e.g.,][]{gerakines95a,kerkhof99}, the IR band strengths of most
molecules show various degrees of dependence upon environmental
factors including ice temperature, overall ice compostion, or the
ice's crystalline phase.

The method by which the column densities of interstellar ices are
determined from their IR absorption features derives from the
Beer-Lambert Law. The fraction of light transmitted through an
interstellar cloud at wavenumber $\nyu$ is given by $T_{\nyu}\ =
\exp(-\tau_{\nyu})$, where the optical depth ($\tau_{\nyu}$) may be
expressed as the Beer-Lambert Law in the form:
\begin{equation}
\label{opticaldepth}
\tau_{\nyu}\ = \sigma_{\nyu}\ N,
\end{equation}
where $N$ is the column density of the absorbing molecule (in
\ns\percmsq), and $\sigma_{\nyu}$ is the absorption cross-section (in
\ns\cmsq) at wavenumber \nyu\ (in \ns\wvn).  Since $N$ is a constant
with respect to $\nyu$, equation~(\ref{opticaldepth}) can be integrated
over wavenumber and re-arranged in the form
\begin{equation}
\label{columndensity}
N = \frac{\int_i \tau_{\nyu}\ d\nyu}{A_i},
\end{equation}
where
\begin{equation}
\label{strength}
A_i = \int_i \sigma_{\nyu}\ d\nyu,
\end{equation}
and the integrals in equations~(\ref{columndensity})
and~(\ref{strength}) are taken over the range of the studied feature
$i$.  The parameter $A_i$ has many names in the literature, including
the ``band strength,'' ``integrated cross-section,'' ``absolute
absorption intensity,'' or simply ``$A$-value'' of the IR absorption
feature $i$.  As expressed above and as is customary in the
astrophysical literature, band strengths possess units of
\ns\cmpmolec, although the physical chemistry literature prefers units
such as km\ mmol$^{-1}$.  The IR absorption features of most molecules
of astrophysical interest have band strengths in the range from about
$10^{-20}$ to $10^{-16}$\cmpmolec\ (\citealt{hudgins93};
\citealt{gerakines95a}; \citealt{kerkhof99}).

Mid-IR absorption features of molecules are the result of their
fundamental vibrational modes, combinations of the fundamental modes,
and their lowest-lying overtones.  Near-IR absorptions are generally
not due to fundamental vibrations but to higher overtones or
combinations of fundamentals and their overtones.  As a result, the
near-IR features of a certain molecule are usually much weaker than
those in the mid-IR.  This is demonstrated in
Figure~\ref{wholespectra}, where the combined near- and mid-IR spectra
of \com, \cod, \suboxide, \methane, \water, \methanol, and \ammonia\
are shown.  Note that in each case, the near-IR portion of the
spectrum (\nyu\ = 10000-4000\wvn; \wav\ = 1.0-2.5\mic) has been
magnified significantly -- from 5 to 100 times -- in order to display
the near- and mid-IR features on similar scales.  Laboratory and
observational studies typically use the much stronger mid-IR
absorption features to make identifications of molecules in
astrophysical ices, due to the fact that they are more sensitive
tracers of molecular abundances.  However, the mid-IR is not always
the best spectral region in which to make astronomical
observations. This is the case for reflectance studies within the
Solar System, since the Solar spectrum peaks in the visible range and
the Sun emits a much higher flux of near-IR photons than mid-IR
photons.

In our laboratory, we have the ability to study the composite near-IR
and mid-IR spectrum, from~10000 to~400\wvn\ (1.0 to~25\mic), of our
ice samples.  This allows us to obtain spectra of the same ice sample
in both the near- and mid-IR ranges almost simultaneously, and thus
spectral characteristics in both regions may be studied for any
material of interest.  On-going work from our laboratory
\citep{cook02,richey04} has been focused on measuring the near-IR
spectra of thick ($\sim$50\mic) ices that have been highly processed
by UV photons.

In this study, we present the near-IR band strengths of \com, \cod,
\suboxide, \methane, \water, \methanol, and \ammonia.  \water,
\methanol, \com, \cod, and \methane\ are confirmed components of
interstellar icy grain mantles, with \water\ dominating most
interstellar ice environments \citep[e.g.,][]{whittet03}.  These are
also ice species found in many icy planetary bodies, including comets
and icy moons such as Europa or Triton \citep[e.g.,][]{roush01}.
Extensive searches for the absorption features of \ammonia\ ice in
interstellar and planetary environments have been performed, with some
varied evidence for their presence (e.g., \citealt{gibb01};
\citealt{gurtler02}; \citealt{taban03}).  The carbon suboxide molecule
(\suboxide) is of interest to both interstellar and planetary
astronomers because it is formed from solid \com\ in environments
affected by energetic processing such as particle irradiation by
cosmic rays or magnetspheric particles, or by ultraviolet photolysis
(\citealt*{huntress91}; \citealt{brucato97}; \citealt{gerakines01a}).

The near-IR band strengths of these molecules were obtained in pure
ice samples at 10\K\ by correlating the growths of the near-IR features
with those in the mid IR over the course of long, slow deposits of
these materials from the gas phase. Future work (currently in the
preliminary phase) will include studies of near- and mid-IR
reflectance spectra as well as the calculations of ice optical
constants in these spectral ranges.

\section{Experimental Methods}

The ice creation methods and the experimental system used in this
study are very similar to those in previous studies by, e.g.,
\citet{hudgins93}, \citet{gerakines95a}, \citet{kerkhof99},
\citet{gerakines00a}, and \citet{gerakines01a}.

Gases are prepared inside a vacuum manifold and vapor-condensed onto
an infrared transmitting substrate (CsI) mounted in the vacuum chamber
($P \approx$ 3\ee{-6}\mmhg\ at room temperature).  The substrate is
cooled by a closed-cycle helium refrigerator (Air Products) to a
temperature of 10\K.  The temperature of the substrate is continuously
monitored by a chromel-Au thermocouple and is adjustable by a
resistive heater element up to room temperature. The chamber that
houses the substrate is accessible to laboratory instruments through
four ports. Two of these ports are have windows composed of KBr,
allowing the transmission of the infrared beam of the spectrometer.
One of the ports contains a window composed of MgF$_2$ to enable
ultraviolet photolysis of the ice samples (this option was not used
for the experiments described in this paper), while the fourth port
has a glass window and is used for visual monitoring.

The sample chamber is positioned within the sample compartment of an
FTIR spectrometer (ThermoMattson Infinity Gold) so that the
spectrometer's IR beam passes through the KBr windows and CsI
substrate.  The spectrometer system is capable of automatic switching
between the appropriate configurations of sources, detectors, and beam
splitters that allow the acquistion of spectra in either the near IR
or mid IR without breaking the dry air purge.  This feature allows the
measurement of spectra in each of these two spectral regions for the
same ice sample in a single experiment.  Near-IR spectra were obtained
over the range from 10000 to 3500\wvn\ (\wav\ = 1.0 to 2.9\mic) and
mid-IR spectra 4500 to 400\wvn\ (\wav\ = 2.2 to 25\mic).  Some overlap
between these ranges allows for cross-calibration of common features
as well as for the concatenation of spectra.  The spectra of \water,
\ammonia, \methanol, \cod, and \suboxide\ were obtained with a
resolution of 4\wvn. The IR spectra of \methane\ and \com\ were taken
at a resolution of 1\wvn.  Absorbance spectra ($\alpha_{\nyu}$) were
calculated from the transmission data ($T_{\nyu}$) by using the
relationship $\alpha_{\nyu} = -\log_{10}{T_{\nyu}}$ (note that
$\tau_{\nyu} = \alpha_{\nyu}\ln{10})$.

A bulb containing the gas to be condensed was prepared on a separate
vacuum manifold and then connected to the high-vacuum system by a
narrow tube through a needle valve, which controlled the gas flow into
the high-vacuum system. The tube is positioned to release the gases
just in front of the cold substrate window.  Gases were deposited at a
rate of about 1-5\mic\per{hr}.  Most deposits spanned several days in
order to create ices thick enough to display detectable near-IR
absorptions.  No significant contamination by background gases in the
vacuum system (dominated by \water) was observed, supported by the
lack of bulk \water\ features in the IR spectra of the non-\water\
samples.  However some small amounts of \water\ may be isolated
within the \com\ and \cod\ samples, as indicated by sharp features
near 3400 and 1600\wvn\ (\wav\ = 2.9 and 6.3\mic) in the mid-IR
spectrum shown in Figure~\ref{wholespectra}.  As indicated by
\citet{gerakines95a}, the band strengths of \com\ and \cod\ are not
significantly affectd by small relative amounts of \water\ in an ice
sample.

It is important to note that ice samples created in this manner at
10\K\ are expected to be amorphous rather than crystalline.  The
amorphous nature of our samples was supported by the IR spectra that
were recorded during each sample deposit (see
Figure~\ref{wholespectra}).

The gases used and their purities are as follows: \com\ (Matheson,
99.9\percent), \cod\ (Matheson, 99.8\percent), \suboxide\
(synthesized), \methane\ (Matheson, 99+\percent), \water\ (distilled
by freeze-thaw cycles under vacuum), \methanol\ (distilled by
freeze-thaw cycles under vacuum), and \ammonia\ (Matheson,
99+\percent).  \water\ and \methanol\ were purified by freezing with
liquid N$_2$ under vacuum and pumping away the more volatile gases
while thawing.  Carbon suboxide was prepared as described in detail by
\cite{gerakines01a}-- who followed the method of \cite{miller64} to
dehydrate malonic acid in the presence of phosphorus pentoxide
(P$_2$O$_5$; a dessicant).  To summarize-- the gases produced by
heating malonic acid to 413\K\ (\water, \cod, acetic acid, and
\suboxide) were collected in a liquid N$_2$ trap and the \suboxide\
was distilled by thermal transfer into a new bulb in the vacuum
manifold.

\section{Results}

The spectra of pure \com, \cod, \suboxide, \methane, \water,
\methanol, and \ammonia\ were collected in the near-IR region from
10000-3500\wvn\ (1.0-2.9\mic) and the mid-IR region from 4500-400\wvn\
(2.2-25\mic) during the slow growth of films at 10\K.
Figure~\ref{wholespectra} displays representative concatenated spectra
(10000-400\wvn; 1.0-25\mic) of these ice samples.  The near-IR regions in
Figure~\ref{wholespectra} have been magnified for direct comparison to
the mid-IR region.  Band strengths were measured by observing the rate
of growth of a near-IR feature in relation to that of a stronger
feature in the mid-IR, as described below.  The mid-IR features used
and their band strengths are listed in Table~\ref{mid-ir}.  The
near-IR features observed in the spectra and their calculated band
strengths are listed in Table~\ref{results_table}.
Figures~\ref{co_plot}-\ref{nh3_plot} display the selected near-IR
regions of the studied spectra and the curves of growth of the near-IR
features.

The integrated absorbance of each feature (in units of \ns\wvn) was
measured by integrating between baseline points on either side of the
feature, assuming a linear shape to the baseline underneath each
feature.  The error in the feature areas was estimated by integrating
across the same limits when no feature was apparent above the noise
level in the infrared spectrum.  For narrow features ($\Delta$\nyu\
$\lesssim$ 10\wvn), the errors were found to be about 0.005~to
0.01\wvn, and larger for wider features (in direct proportion to the
width). For this reason, any areas with measured values below about
0.008\wvn\ were omitted from the fitting procedure (these points are
plotted as open symbols in Figures~\ref{co_plot}-\ref{nh3_plot}).

To determine the relative strengths of two absorption features for a
single molecule, we have examined the relationship between the areas
of these bands as the ice is grown.  In principle, one could merely
measure a single ice spectrum and determine the relative strengths of
any two features by taking the ratio of their areas.  However, since
the near-IR features are much weaker than those in the mid-IR
(typically, 10-100 times smaller; see Figure~\ref{wholespectra}), errors
in their measured areas are relatively much higher.  By monitoring the
relationship between two given absorption features over the course of
a slow deposit, the systematic errors involved in measuring the areas
of the near-IR features are signifcantly reduced.

For optically thin ice absorptions, we expect that depositing a
certain number of mol\-e\-cules will cause a linear increase the areas of
both features, where each increase is proportional to that feature's
band strength-- see equation~\ref{columndensity}.  Hence, the initial
trend in a plot of feature area vs. feature area should be a straight
line, whose slope corresponds to the ratio of the band strengths.
Multiplying this ratio by the known band strength of the mid-IR
feature (Table~\ref{mid-ir}), the near-IR band strength is obtained.
The experimental errors in near-IR band strengths were obtained by
multiplying the standard deviations in the best-fitting slopes by the
mid-IR band strength. Values resulting from the trends displayed in
Figures~\ref{co_plot}-\ref{nh3_plot} are listed in
Table~\ref{results_table}.

In the cases of \com, \cod, and \methane, the scaling process was
complicated by the fact that the fundamental features are extremely
sharp and become saturated very quickly during the deposit.  For \com\
(Figure~\ref{co_plot}), the absorption due to its fundamental
vibration at 2137\wvn\ (4.679\mic) becomes too large to be useful
after only short deposition times.  As a result, we have used the
absorption feature of \thcom\ at 2092\wvn\ (4.780\mic) to scale the
\twcom\ and \thcom\ near-IR features.  The strength of
1.5\ee{-19}~cm~per~\twcom\ molecule was used for the 2092\wvn\
(4.780\mic) feature, taking the band strength of the \thcom\ feature
from \cite{gerakines95a} and scaling by the terrestrial ratio of
$^{12}$C/$^{13}$C of 87.  For \cod, the combination mode at 3708\wvn\
(2.697\mic) was used to scale the near-IR features.  For \methane\
(Figure~\ref{ch4_plot}), the absorptions due to fundamental modes at
3009 and 1306\wvn\ (3.323 and 7.657\mic) become too strong to be
useful after only short timescales as well.  Because of this, the
near-IR features were scaled by the absorption feature located at
2815\wvn\ (3.552\mic), which is due to its $\nu_2 + \nu_4$ vibration
mode.  In order to to this, we first determined the band strength for
the 2815\wvn\ (3.552\mic) feature to be $A$ =
(1.9$\pm$0.1)\ee{-18}\cmpmolec\ by scaling it by the $\nu_2$
fundamental mode at 1306\wvn\ (\wav\ = 7.657\mic; $A$ =
7.0\ee{-18}\cmpmolec; \citealt{kerkhof99}) in the spectrum of a thin
sample.  The 8405\wvn\ (1.190\mic) feature of \methane\ displayed a
peculiar growth curve, with no discernible linear trend.  Hence, no
band strength is listed for this feature in Table~\ref{results_table}.

As observed in Figures~\ref{co_plot}-\ref{nh3_plot}, the trends are
initially linear for the ices we have studied.  In some cases, the
mid-IR feature used for the $x$-axis becomes so large that its curve
of growth becomes non-linear (the change in feature area no longer
responds linearly to the increase in the number of molecules).  In
these cases, the non-linear portions of the data set were omitted from
our fitting process (omitted data points are plotted as open symbols
in Figures~\ref{co_plot}-\ref{nh3_plot}).  It may be interesting to
note that the non-linear parts of the curve are not always identical
from experiment to experiment.  This is especially clear in the two
separate deposits of \suboxide\ (Fig.~\ref{c3o2_plot}), \water\
(Fig.~\ref{h2o_plot}), and \ammonia\ (Fig.~\ref{nh3_plot}).  This may
suggest that ice deposition rates or other characteristics of a single
experiment can significantly alter the physical properties of the ice
under study, especially when ices are thicker than about 10\mic.  This
particular issue has been discussed in detail previously by
\citet{quirico97}.

\section{Comparison to previous studies}

\cite{taban03} have published values for some of the same near-IR band
strengths of pure \water, \ammonia, and \methanol\ ices at low
temperatures.  In general, we find our values to be in excellent
agreement with theirs.  They quote $A$ = 1.1\ee{-18}\cmpmolec\ for the
\water\ feature near 5000\wvn\ (2.0\mic), which is in full agreement
with our experimentally determined value of
$(1.2\pm0.1)$\ee{-18}\cmpmolec.  For \ammonia, they find a value of
$A$ = 9.7\ee{-19}\cmpmolec\ for the band near 4478\wvn\ (2.233\mic),
for which we find $A$ = $(8.7\pm0.3)$\ee{-19}\cmpmolec.  For
\methanol, they measure $A$ = $5.9$\ee{-19}\cmpmolec\ for the feature
near 4395\wvn\ (2.275\mic), which is also in good agreement with our
measured value of $(8.7\pm0.7)$\ee{-19}\cmpmolec\ for the 4400\wvn\
(2.273\mic) feature observed here.

\cite{quirico97} published absorption coefficients (the fraction of
intensity absorbed per unit thickness of the sample; in \wvn) of the
near-IR features of pure \com, \cod, and \methane\ for use in
planetary ice studies.  The absorption coefficient allows one to
calculate the thickness of a sample from the absorbance value at the
peak of a feature.  It is not necessarily straightforward to compare
band strengths with absorption coefficients for ice samples prepared
in separate laboratories using different techniques, since one must
take into account the densities of the samples in order to connect
column densities to thicknesses.  Ice samples in the \citet{quirico97}
study were created in a closed cell and not by vapor deposit.  For the
sake of a comparison to our study, one must make the rather unsafe
assumption that the ice densities and spectral profiles of our samples
are identical to theirs (they most likely are not identical).  In this
case, the band strengths of any two features for a given molecule
should scale in the same manner as their absorption coefficients.
Taking the ratios of band strengths from Table~\ref{results_table} for
pairs of \com, \cod, and \methane\ features and comparing them to the
ratios of their reported absorption coefficients \cite{quirico97}, we
find that some agree quite well (to within a few percent), but most
agree only to within a factor of 2 or so.

\cite{taban03} find that the near-IR spectrum of the line of sight
toward the high-mass protostar W33A is consistent with the known
abundances of \water, \methanol, and \com\ as derived from mid-IR
data.  They claim the detection of the near-IR features of \methanol\
near 4400\wvn\ (2.273\mic) with an optical depth of about 0.014.

Based on our laboratory data, one should be able to predict the
optical depths of near-IR ice absorptions that could be investigated
in various astrophysical objects.  Using the widths and strengths for
the two strongest near-IR features of our samples, we have estimated
the their optical depths for some well-studied lines of sight in the
ISM \citep[e.g.,][]{gerakines99,gibb00}.  These estimates are listed
in Table~\ref{optical_depths}.  Although pure ices at 10\K\ may not
reflect the most realistic cases for these objects, previous work
\citep{gerakines95a,kerkhof99} does suggest that the band strengths of
most molecules studied to date do not vary by more than a factor of 2
or so according to composition or to temperature. It should be noted
that the strengths of a certain molecule's absorption features do vary
by large amounts when different crystalline states are considered, but
interstellar ices are presumed to be amorphous
\citep[e.g.,][]{whittet03}.

\section{Summary and Future Work}

In this paper, we have shown that the near-IR band strengths
(or ``absorption intensities'') for molecules of interest to both
interstellar and planetary astronomers may be determined through
correlations to their better-known mid-IR characteristics.  By
correlating the growth of near-IR and mid-IR absorption features for
molecules at low temperature, we have calculated the absorption
strengths for the near-IR features of \com, \cod, \suboxide, \methane,
\water, \methanol, and \ammonia.  These strengths may be used to
determine the column densities of these molecules in the interstellar
dense cloud or other environments from observed transmission data.

This is the first paper in a series of near- and mid-IR correlation
studies of ice samples of astrophysical interest.  Future work
currently in preparation in our laboratory will involve the
calculation of ice optical constants for use in particle scattering
models as well as reflectance studies for use in the direct
interpretation of planetary observations of reflected sunlight.  We
will also investigate the effects, if any, of ice composition on the
near-IR band strengths as well as the differences bewteen the near-IR
band strengths of crystalline and amorphous ices.

\acknowledgments

PAG gratefully acknowledges laboratory start-up funds from the
University of Alabama at Birmingham, financial support through NASA
grant number NNG04GA63A, and many conversations with Marla Moore and
Reggie Hudson, who kindly provided comments on an early version of
this manuscript.

\clearpage
\begin{deluxetable}{lcccc}
\tablecolumns{5}
\tablecaption{Mid-Infrared Features Used in Strength Determinations\label{mid-ir}}
\tablehead{\colhead{Molecule} & \colhead{$\nyu$ [\ns\wvn]} & \colhead{\wav\ [\ns\mic]} & \colhead{$A$ [\ns\cmpmolec]} & Reference}
\startdata
\com      & 2092 & 4.780 & 1.5\ee{-19}~\tablenotemark{a} & 1 \\
\cod      & 3708 & 2.697 & 1.4\ee{-18}                  & 1 \\
\suboxide & 3744 & 2.671 & 3.8\ee{-18}                  & 2 \\
\methane  & 1306 & 7.657 & 7.0\ee{-18}                  & 3 \\
          & 2815 & 3.552 & ($1.9\pm0.1$)\ee{-18}        & 4 \\
\water    & 1670 & 5.988 & 1.2\ee{-17}                  & 1 \\
\methanol & 2830 & 3.534 & 7.6\ee{-18}                  & 5 \\
\ammonia  & 1070 & 9.346 & 1.2\ee{-17}                  & 3 \\
\enddata
\tablenotetext{a}{Although this is a feature of \thcom, its band strength is expressed in units of cm per \twcom\ molecule.}
\tablecomments{(1) \cite{gerakines95a};
               (2) \cite{gerakines01a};
               (3) \cite{kerkhof99};
               (4) determined by scaling the 1306\wvn\ (7.657\mic) feature (see text);
               (5) \cite{dhendecourt86a}.}
\end{deluxetable}

\clearpage

\begin{deluxetable}{cccccc}
\tablecolumns{5}
\tablecaption{Near-IR Features and Band Strengths Measured at 10\K \label{results_table}}
\tablehead{
\multicolumn{2}{c}{Peak Position}         & \colhead{FWHM, $\Delta$\nyu} & \colhead{}           & \colhead{Strength, $A$}  & \colhead{}      \\
\cline{1-2}
\colhead{\nyu\ [\ns\wvn]} & \colhead{$\lambda$ [\ns\mic]} & \colhead{[\ns\wvn]}   & \colhead{Vibration Mode} & \colhead{[\ns\cmpmolec]} & \colhead{Notes} }
\startdata
\sidehead{\com\tablenotemark{a}}
4159 & 2.404 & 3 & $2\nu$ \thcom\     & $(2.1\pm0.1)\ee{-21}$ & 1 \\
4252 & 2.352 & 3 & $2\nu$ \twcom\     & $(1.6\pm0.1)\ee{-19}$ & 1 \\
6338 & 1.578 & 4 & $3\nu$ \twcom\     & $(1.5\pm0.1)\ee{-21}$ & 1 \\
8504 & 1.176 & 8 & ?                  & $(3.4\pm0.2)\ee{-21}$ &     \\
\sidehead{\cod}
4832 & 2.070 &  6 & $4\nu_2 + \nu_3$          & $(7.8\pm0.5)\ee{-21}$ & 1 \\
4971 & 2.012 &  7 & $\nu_1 + 2\nu_2 + \nu_3$  & $(5.4\pm0.2)\ee{-20}$ & 1 \\
5087 & 1.966 &  7 & $2\nu_1 + \nu_3$          & $(2.8\pm0.1)\ee{-20}$ & 1 \\
6214 & 1.609 &  7 & $\nu_1 + 4\nu_2 + \nu_3$  & $(2.6\pm1.8)\ee{-22}$ & 1 \\
6341 & 1.577 &  7 & $2\nu_1 + 2\nu_2 + \nu_3$ & $(8.6\pm1.1)\ee{-22}$ & 1 \\
6972 & 1.434 &  8 & $3\nu_3$                  & $(2.6\pm0.3)\ee{-21}$ & 1 \\
9944 & 1.006 & 12 & $4\nu_3 + \nu_2$ (?)      & $(5.6\pm1.9)\ee{-21}$ & 1    \\
\sidehead{\suboxide}
4370  & 2.288 & 78 & $2\nu_3$ (?)                & $(1.7\pm0.2)\ee{-19}$ &  \\
4570  & 2.188 & 37 & $\nu_2 + \nu_3 + \nu_4$ (?) & $(1.7\pm0.2)\ee{-20}$ &  \\
5270  & 1.898 & 48 & $\nu_2 + 2\nu_3$ (?)        & $(5.0\pm0.9)\ee{-20}$ &  \\
\sidehead{\methane}
4115 & 2.430 & 14 & $\nu_2 + 2\nu_4$                        & $(7.2\pm0.1)\ee{-20}$ & 1 \\
4202 & 2.380 &  9 & $\nu_1 + \nu_4$                         & $(1.6\pm0.1)\ee{-18}$ & 1 \\
4300 & 2.326 &  7 & $\nu_3 + \nu_4$                         & $(3.4\pm0.1)\ee{-18}$ & 1 \\
4528 & 2.208 &  9 & $\nu_2 + \nu_3$                         & $(4.5\pm0.1)\ee{-19}$ & 1 \\
5565 & 1.797 &  5 & $\nu_3 + 2\nu_4$                        & $(6.8\pm0.1)\ee{-20}$ & 1 \\
5800 & 1.724 &  8 & $\nu_2 + \nu_3 + \nu_4$                 & $(1.3\pm0.1)\ee{-19}$ & 1 \\
5987 & 1.670 &  7 & $2\nu_3$                                & $(3.8\pm0.1)\ee{-19}$ & 1 \\
7487 & 1.336 & 11 & $\nu_2 + 2\nu_3$                        & $(5.3\pm0.1)\ee{-20}$ & 1 \\
8405 & 1.190 & 30 & $[\nu_1] + [\nu_1] + [\nu_4] + [\nu_4]$ & \nodata               & 2 \\
8588 & 1.164 & 42 & $2\nu_3 + 2\nu_4$                       & $(5.4\pm0.1)\ee{-20}$ & 1 \\
8780 & 1.139 & 17 & $\nu_2 + 2\nu_3 + \nu_4$                & $(2.0\pm0.1)\ee{-20}$ & 1 \\
9060 & 1.104 & 28 & $3\nu_3$                                & $(2.1\pm0.2)\ee{-20}$ & 3 \\
\sidehead{\water}
5040 & 1.984 & 408 & $\nu_2 + \nu_3$ & $(1.2\pm0.1)\ee{-18}$ &  \\
6684 & 1.496 & 520 & $2\nu_3$        & $(8.8\pm1.0)\ee{-19}$ &  \\
\sidehead{\methanol}
4280 & 2.336 & 34 & ? & $(8.0\pm0.9)\ee{-20}$ & \\
4400 & 2.273 & 72 & ? & $(8.7\pm0.7)\ee{-19}$ & \\
\sidehead{\ammonia}
4474 & 2.235 &  94 & $\nu_1 + \nu_2$ (?)         & $(8.7\pm0.3)\ee{-19}$ & \\
4993 & 2.002 &  68 & $\nu_1 + \nu_4$ (?)         & $(8.1\pm0.4)\ee{-19}$ & \\
6099 & 1.640 & 107 & $\nu_1 + \nu_2 + \nu_4$ (?) & $(2.8\pm0.5)\ee{-20}$ & \\
6515 & 1.535 & 118 & $2\nu_1$ (?)                & $(3.9\pm0.3)\ee{-20}$ & \\
\enddata
\tablenotetext{a}{All \twcom\ and \thcom\ band strengths are expressed in units of cm per \twcom\ molecule.} 
\tablecomments{(1) vibrational mode assignment from \cite{calvani92} and \cite{quirico97};
               (2) combination of dual-phonon modes \citep{calvani92,quirico97};
               (3) vibrational mode assignment from \cite{grundy02}.}
\end{deluxetable}

\clearpage

\begin{deluxetable}{lcccc}
\tablecolumns{5}
\tablecaption{Column Densities\tablenotemark{a} and Predicted Near-IR Optical Depths\tablenotemark{b} of Interstellar Ices\label{optical_depths}}
\tablehead{%
\colhead{} & \colhead{W33A} & \colhead{NGC 7538 IRS9} & \colhead{Elias~16} & \colhead{Sgr A$^*$}}
\startdata
\sidehead{CO}
$N$               & 8.8\ee{17} & 1.2\ee{18} & 6.3\ee{17}  & $<$2\ee{17} \\
$\tau$(2.352\mic) & 0.05       & 0.06      & 0.03       & $<$0.01       \\
\sidehead{\cod}
$N$               & 1.4\ee{18} & 1.7\ee{18} & 4.5\ee{17}  & 1.8\ee{17}  \\ 
$\tau$(1.966\mic) & 0.006      & 0.007      & 0.002       & $<$0.001    \\
$\tau$(2.012\mic) & 0.01       & 0.01      & 0.004       & 0.001        \\
\sidehead{\methane}
$N$              & 1.7\ee{17} & 1.5\ee{17} & \nodata     & 2.6\ee{16}  \\
$\tau$(2.326\mic) & 0.08      & 0.07       & \nodata     & 0.01        \\
$\tau$(2.380\mic) & 0.03      & 0.03       & \nodata     & 0.005       \\
\sidehead{\water}
$N$               & 1.1\ee{19} & 7.5\ee{18} & 2.5\ee{18}  & 1.3\ee{18}  \\
$\tau$(1.496\mic) & 0.02       & 0.01      & 0.004       & 0.002        \\
$\tau$(1.984\mic) & 0.03       & 0.02      & 0.007       & 0.004        \\
\sidehead{\methanol}
$N$               & 2.0\ee{17} & 3.8\ee{17} & $<$8\ee{16} & $<$5\ee{16} \\   
$\tau$(2.273\mic) & 0.02       & 0.005      & $<$0.01     & $<$0.006    \\
$\tau$(2.336\mic) & 0.005      & $<$0.001   & $<$0.002    & $<$0.001    \\
\enddata
\tablenotetext{a}{Column densities ($N$) have units of cm$^{-2}$; from \cite{gibb00}.}
\tablenotetext{b}{Optical depths have been calculated from the listed column densities, using width and strength data listed in Table~\ref{results_table}.} 
\end{deluxetable}

\clearpage

\figcaption{Combined near- and mid-IR spectra (10000-400\wvn;
  1.0-25\mic) of pure samples at 10\K\ for the seven molecules
  studied.  In each case, the near-IR portion of the spectrum
  (10000-4000\wvn; 1.0-2.5\mic) has been magnified by the amount
  indicated in the Figure.  The sinusoidal appearance of the CO
  near-IR spectrum is due to interference of the spectrometer's IR
  beam within the sample. Gas-phase \water\ lines from the imperfect
  spectrometer purge (5540-4100\wvn\ and 7420-7000\wvn) have been
  removed from the near-IR spectra of CO and \methane. The
  10000-7000\wvn\ region has been omitted from the near-IR spectra of
  \suboxide\ and \methanol\ for clarity, since they are dominated by
  noise.
  \label{wholespectra}}

\figcaption{(top) Selected near-IR absorbance spectra of an 18\mic\
  thick \com\ ice at 10\K\ for the regions surrounding the features
  studied; (bottom) Integrated areas (in \ns\wvn) of \com\ near-IR
  features plotted versus the area (in \ns\wvn) of the \thcom\ feature
  at 2092\wvn\ (4.780\mic) during deposition at 10\K.  Solid lines--
  linear fits to the solid symbols.  Data points plotted with empty
  symbols were omitted from the fits (see text).\label{co_plot}}

\figcaption{(top) Selected regions of the near-IR absorbance spectrum
  of an $\sim$14\mic\ thick \cod\ ice at 10\K\ for the regions
  surrounding the features studied; (bottom) Integrated areas (in
  \ns\wvn) of \cod\ near-IR features plotted versus the area (in
  \ns\wvn) of the \cod\ feature at 3708\wvn\ (2.697\mic) during
  deposition at 10\K.  Lines and symbols have the same meaning as in
  Figure~\ref{co_plot}.\label{co2_plot}}

\figcaption{(top) Selected regions of the near-IR absorbance spectrum
  of a thick \suboxide\ sample at 10\K. (bottom) Integrated areas (in
  \ns\wvn) of \suboxide\ near-IR features plotted versus the area (in
  \ns\wvn) of the \suboxide\ feature at 3700\wvn\ (2.703\mic) during
  deposition at 10\K.  Data points plotted as squares were measured
  from a separate deposit experiment from those plotted with circles.
  Otherwise, the lines and symbols have the same meaning as in
  Figure~\ref{co_plot}.\label{c3o2_plot}}

\figcaption{(top) Selected resgions of the near-IR absorbance spectrum
  of a thick \methane\ sample at 10\K.  (bottom) Integrated areas (in
  \ns\wvn) of \methane\ near-IR features plotted versus the area (in
  \ns\wvn) of the \methane\ feature at 2815\wvn\ (3.552\mic) during
  deposition at 10\K.  Lines and symbols have the same meaning as in
  Figure~\ref{co_plot}.\label{ch4_plot}}

\figcaption{(top) Near-IR absorbance spectrum of an $\sim$11\mic\
  thick \water\ ice at 10\K\ in the range of 7500--4000\wvn\
  (1.333--2.5\mic), displaying the features at 6684 and 5040\wvn\
  (1.496 and 1.984\mic); (bottom) Integrated areas (in \ns\wvn) of
  \water\ near-IR features plotted versus the area (in \ns\wvn) of the
  \water\ feature at 1670\wvn\ (5.988\mic) during deposition at 10\K.
  Lines and symbols have the same meaning as in
  Figure~\ref{co_plot}.\label{h2o_plot}}

\figcaption{(top) Near-IR absorbance spectrum of a thick \methanol\
  sample at 10\K; (bottom) Integrated areas (in \ns\wvn) of \methanol\
  near-IR features plotted versus the area (in \ns\wvn) of the
  \methanol\ feature at 2830\wvn\ (3.534\mic) during deposition at
  10\K.  Lines and symbols have the same meaning as in
  Figure~\ref{co_plot}.\label{ch3oh_plot}}

\figcaption{(top) Selected regions of the near-IR absorbance spectrum
  of a thick \ammonia\ sample at 10\K; (bottom) Integrated areas (in
  \ns\wvn) of \ammonia\ near-IR features plotted versus the area (in
  \ns\wvn) of the \ammonia\ feature at 1070\wvn\ (9.346\mic) during
  deposition of \ammonia\ at 10\K.  Data points plotted as squares
  were measured from a separate deposit experiment from those plotted
  with circles.  Otherwise, lines and symbols have the same meaning as
  in Figure~\ref{co_plot}.\label{nh3_plot}}

\clearpage

\begin{figure}
\centerline{FIGURE~\ref{wholespectra}}
\epsscale{1.0}
\plotone{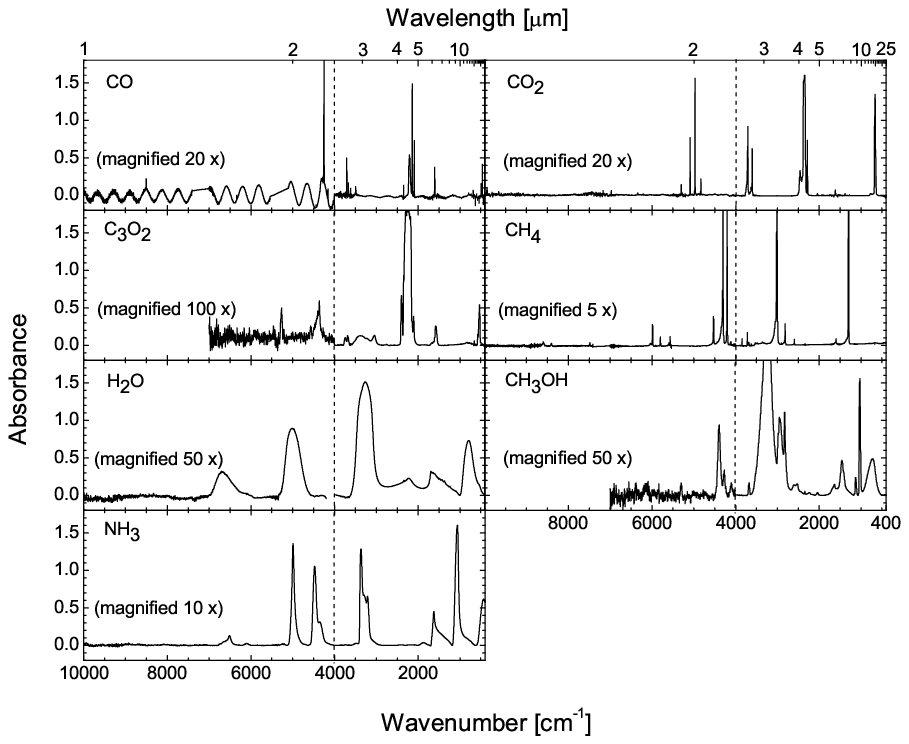}
\end{figure}

\clearpage

\begin{figure}
\centerline{FIGURE~\ref{co_plot}}
\epsscale{0.8}
\plotone{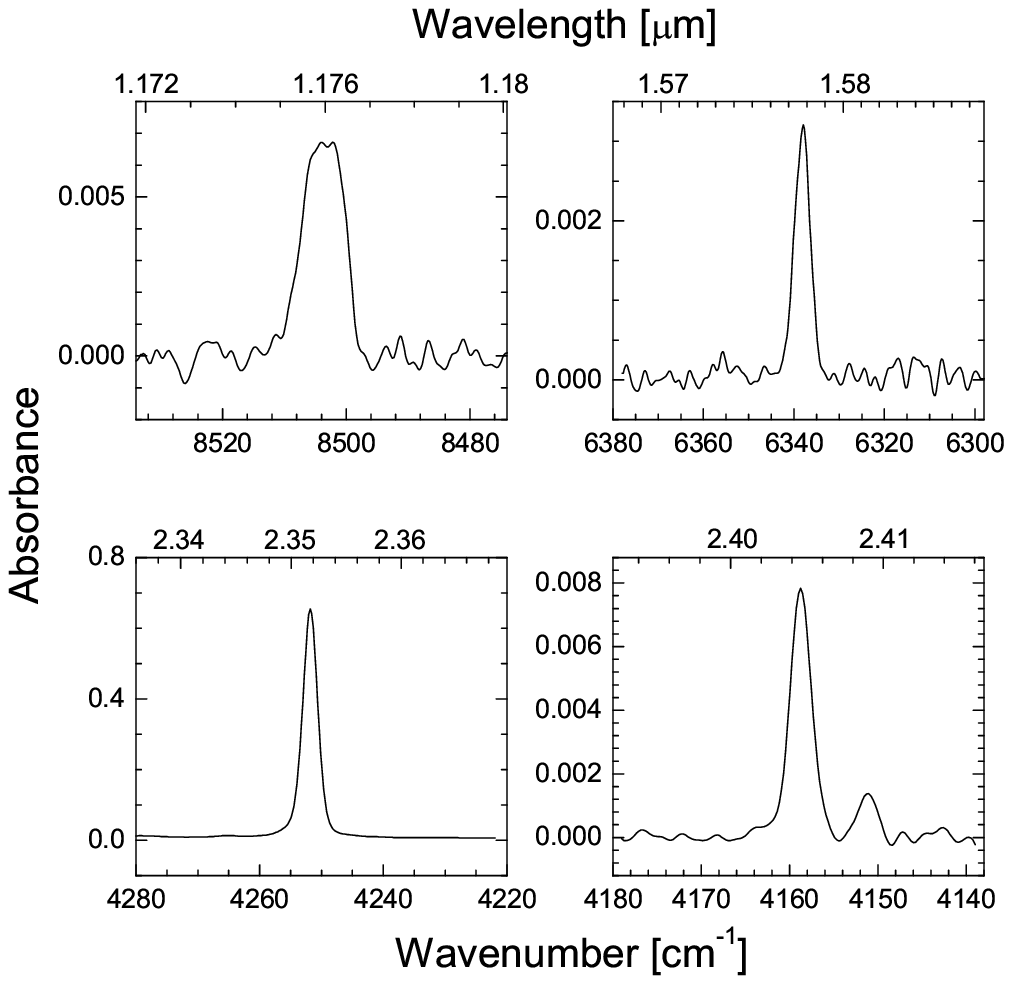}
\plotone{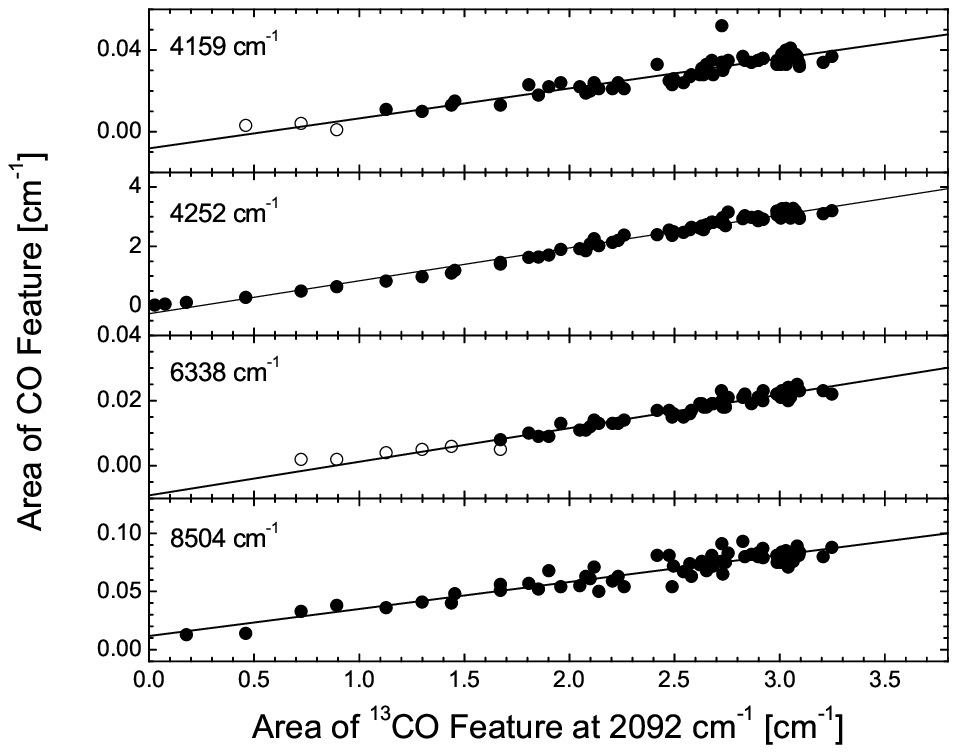}
\epsscale{1.0}
\end{figure}

\clearpage

\begin{figure}
\centerline{FIGURE~\ref{co2_plot}}
\epsscale{0.8}
\plotone{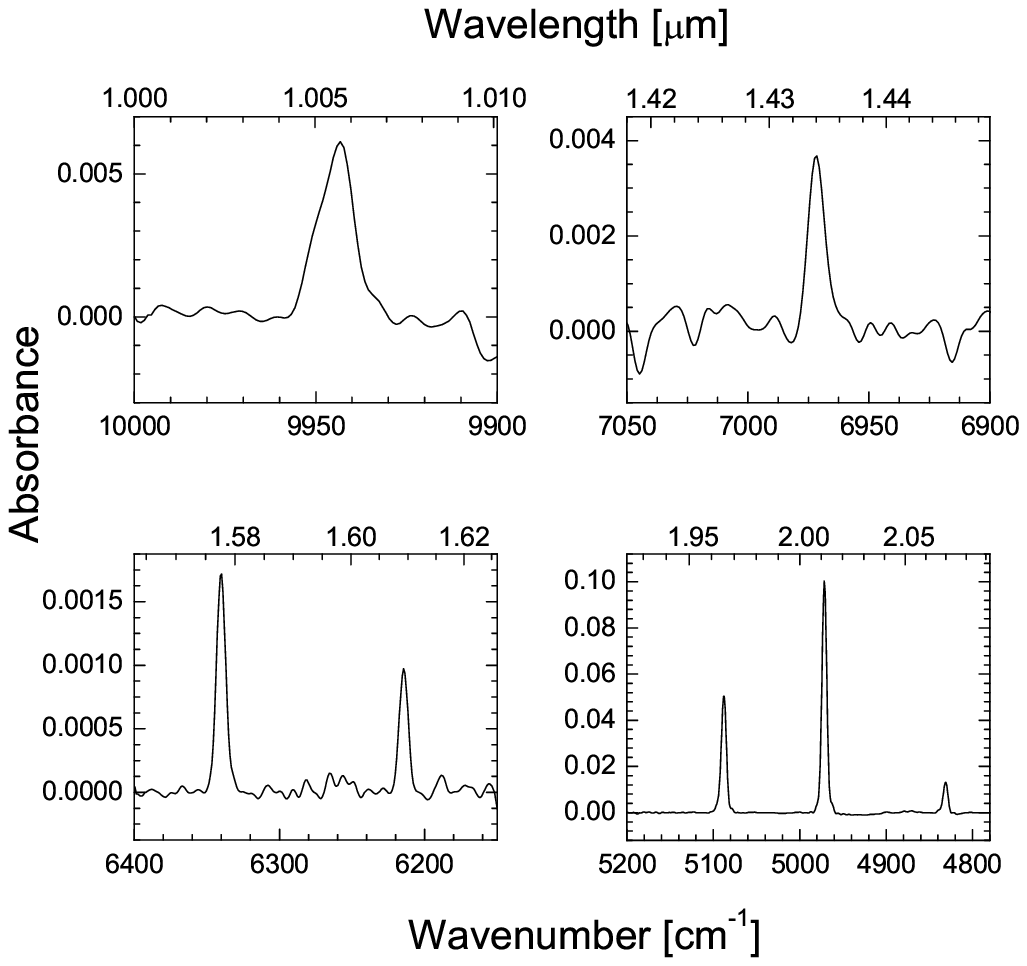}
\plotone{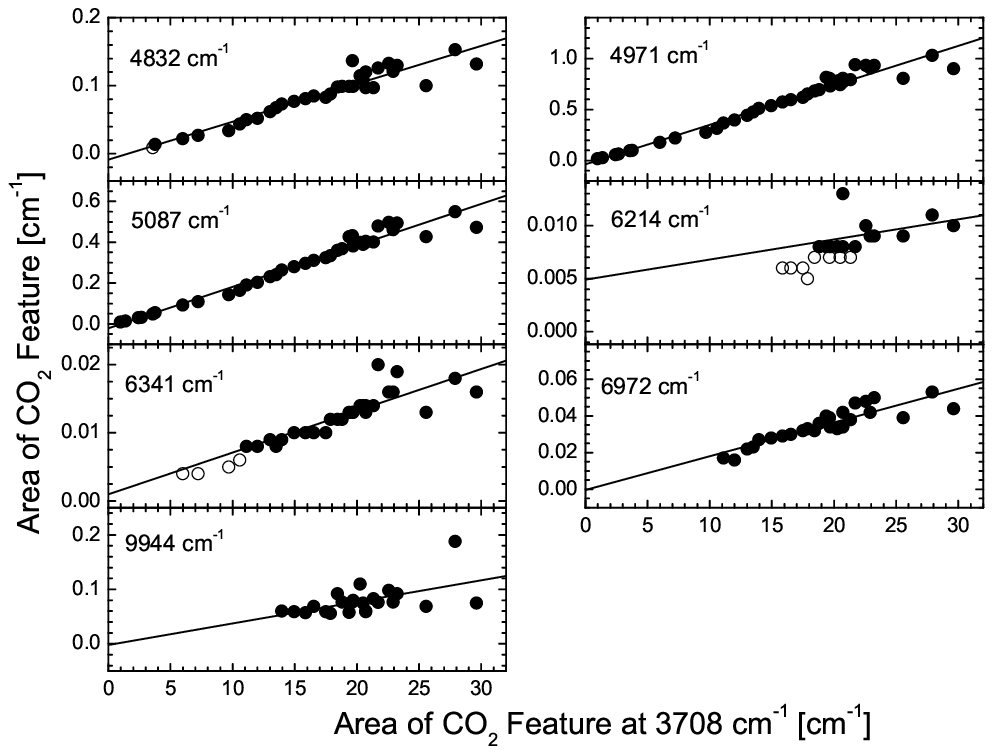}
\epsscale{1.0}
\end{figure}

\clearpage

\begin{figure}
\centerline{FIGURE~\ref{c3o2_plot}}
\epsscale{0.8}
\plotone{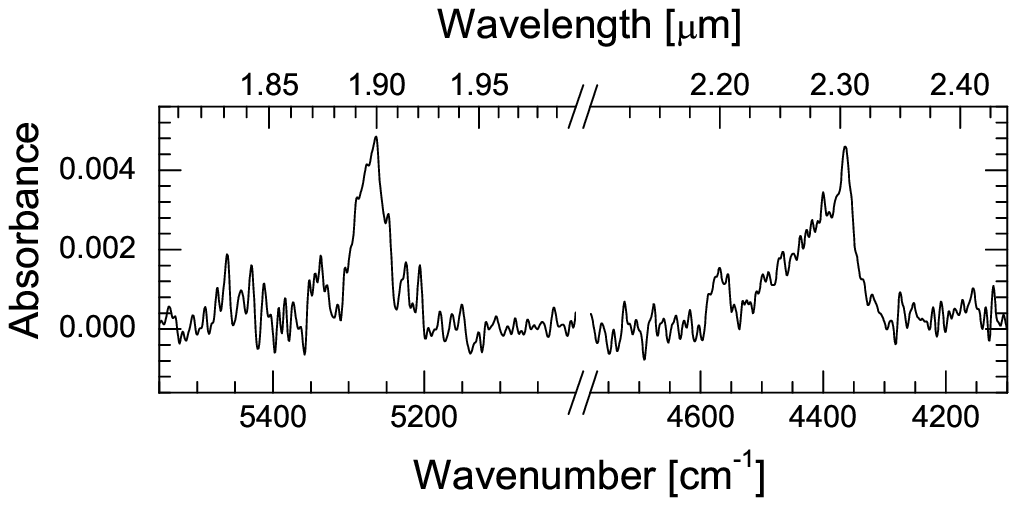}
\plotone{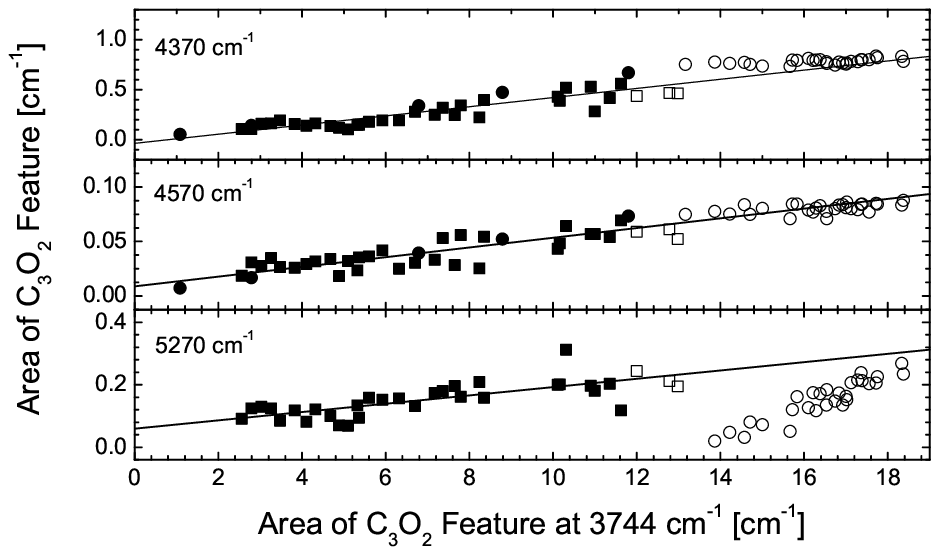}
\epsscale{1.0}
\end{figure}

\clearpage

\begin{figure}
\centerline{FIGURE~\ref{ch4_plot}}
\epsscale{0.8}
\plotone{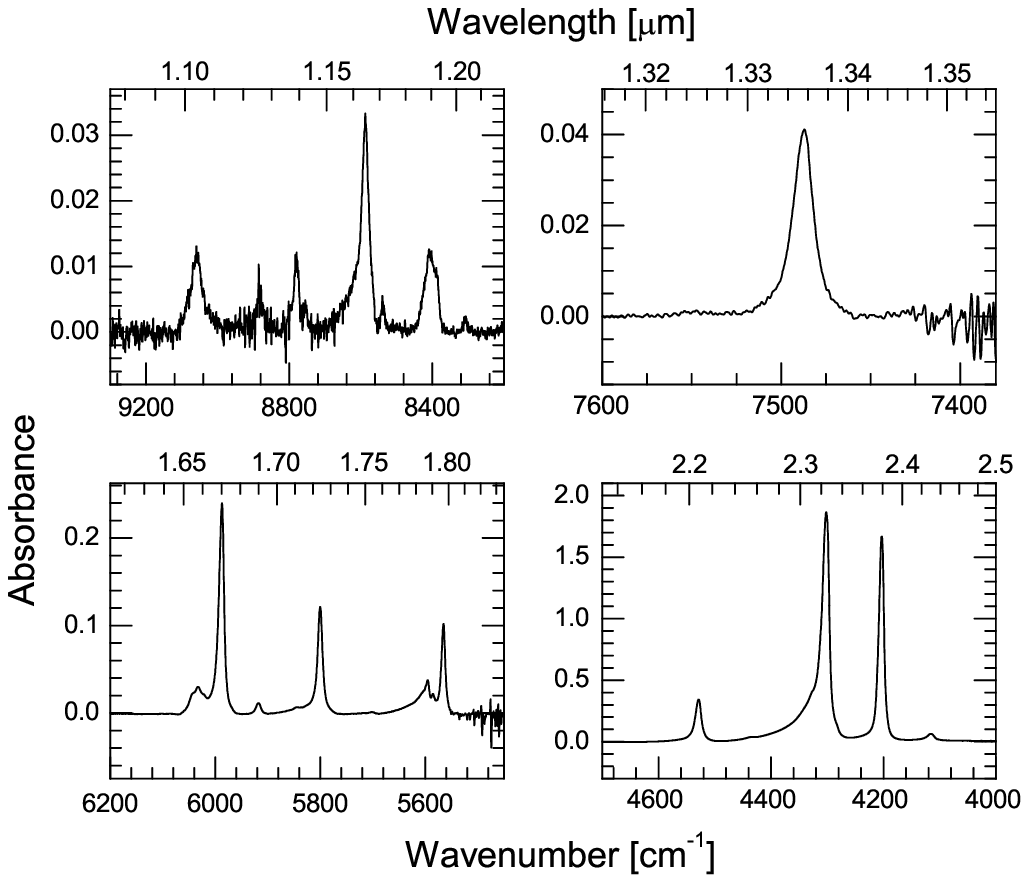}
\plotone{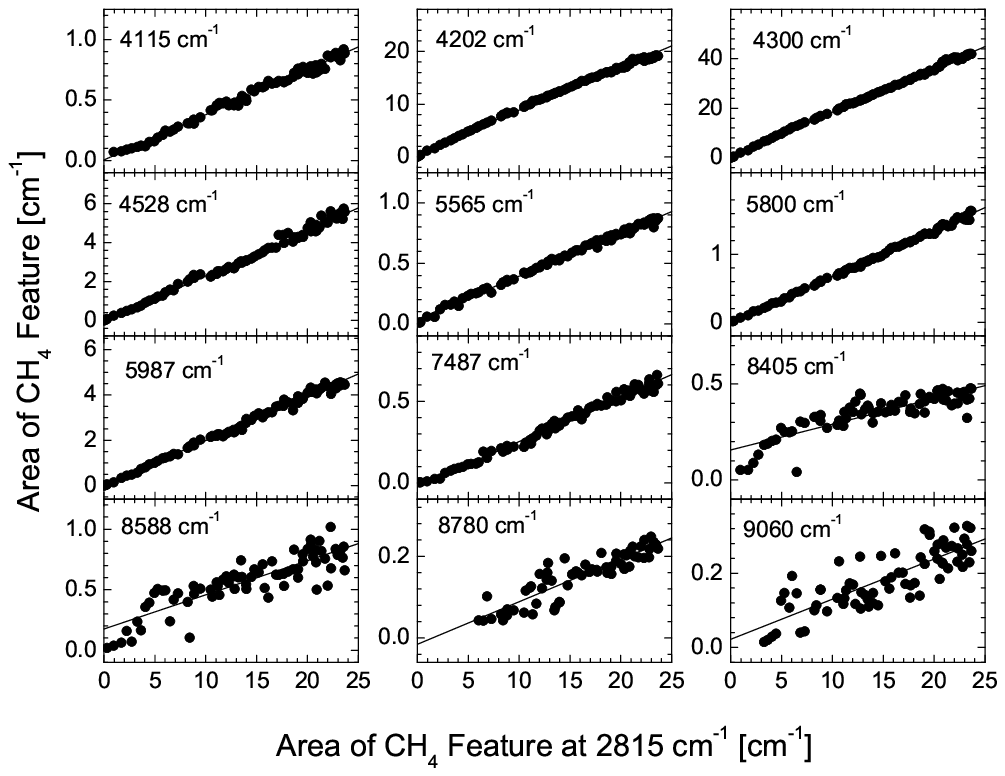}
\epsscale{1.0}
\end{figure}

\clearpage

\begin{figure}
\centerline{FIGURE~\ref{h2o_plot}}
\epsscale{0.8}
\plotone{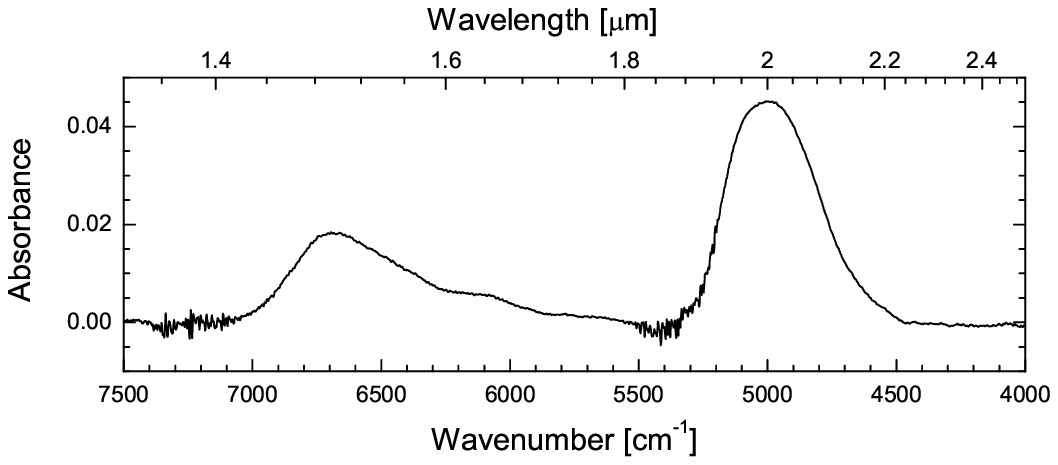}
\plotone{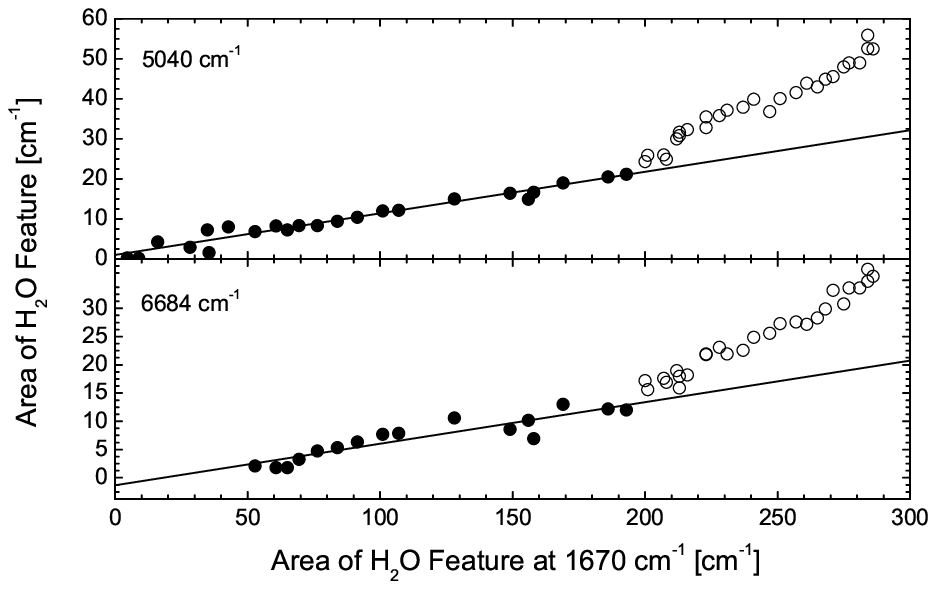}
\epsscale{1.0}
\end{figure}

\clearpage

\begin{figure}
\centerline{FIGURE~\ref{ch3oh_plot}}
\epsscale{0.8}
\plotone{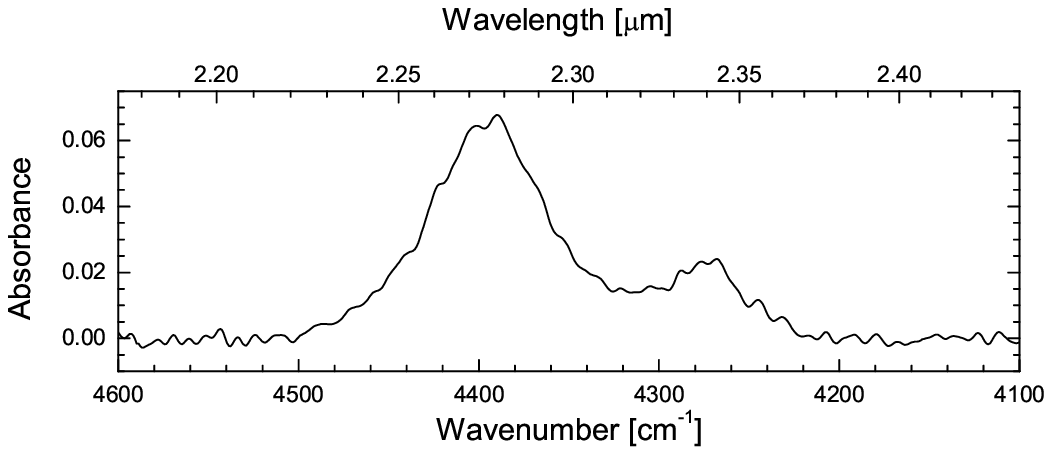}
\plotone{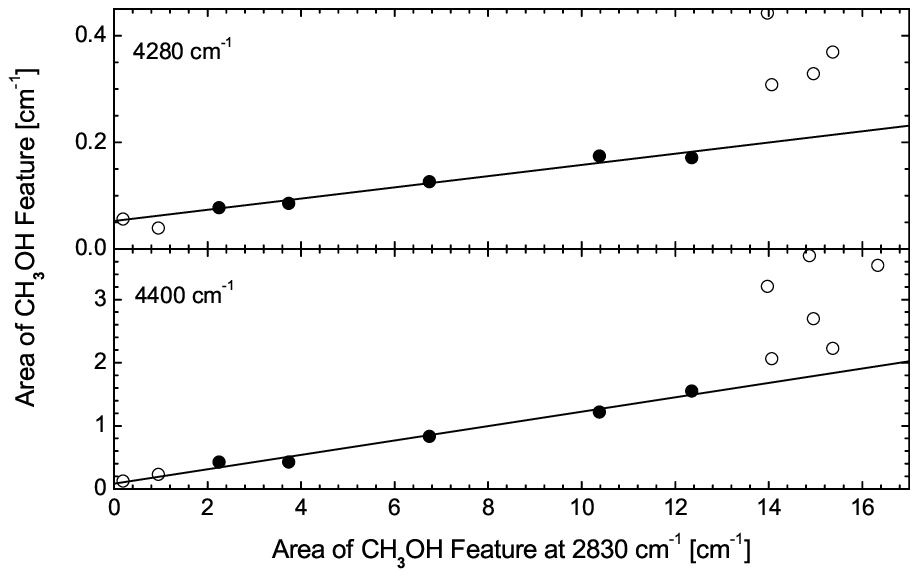}
\epsscale{1.0}
\end{figure}

\clearpage

\begin{figure}
\centerline{FIGURE~\ref{nh3_plot}}
\epsscale{0.8}
\plotone{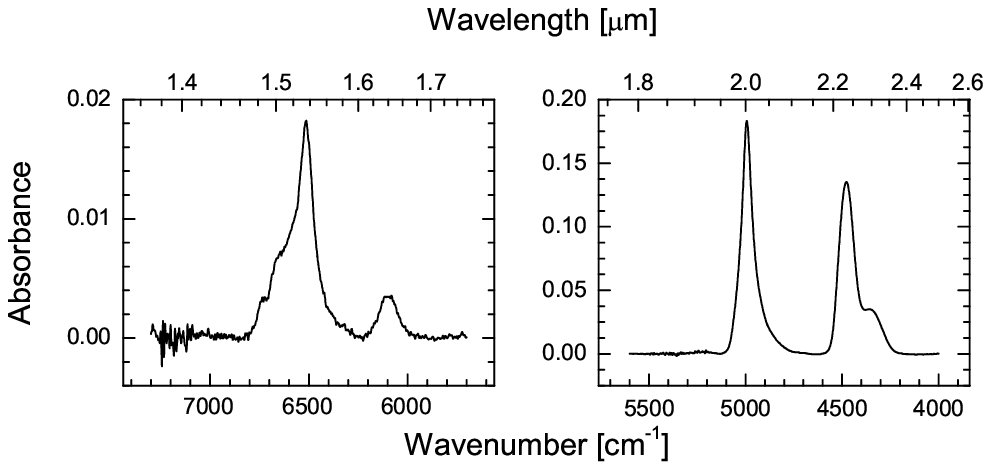}
\plotone{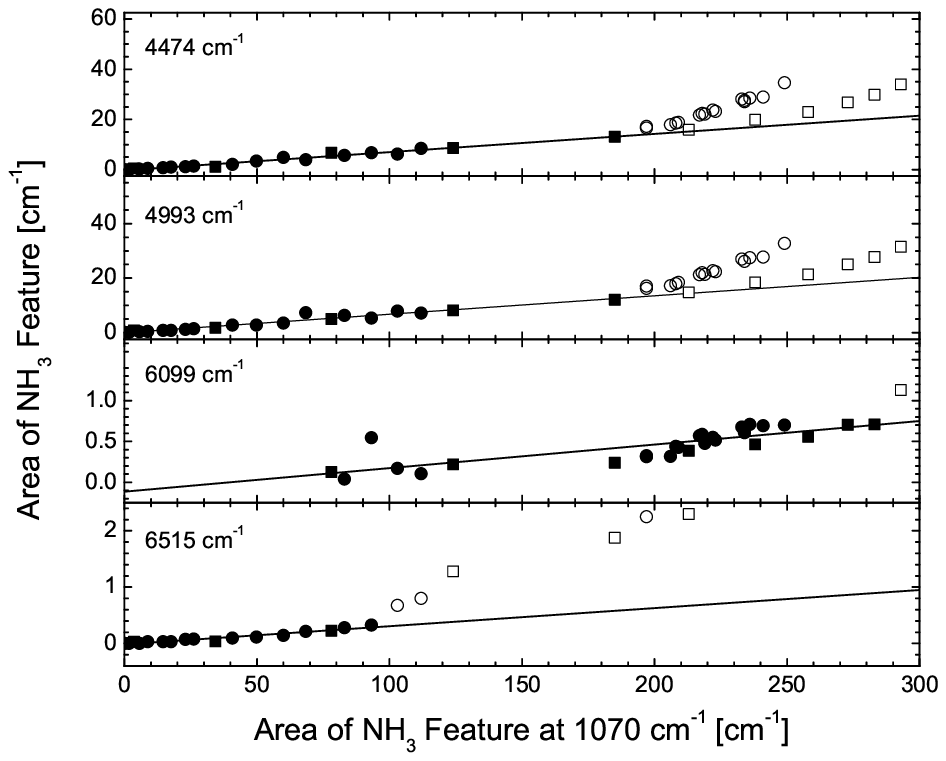}
\epsscale{1.0}
\end{figure}

\end{document}